\documentstyle[11pt,IAU207_pasp,twoside,epsf,psfig]{article}
\def\edcomment#1{\iffalse\marginpar{\raggedright\sl#1\/}\else\relax\fi}
\reversemarginpar

\begin{document}
\title{Kinematics of Globular Clusters in M49 and M87}
\author{Terry Bridges}
\affil{Anglo-Australian Observatory, P.O. Box 296,
Epping, NSW, Australia}

\begin{abstract}
We present recent multi-object spectroscopy of globular clusters in
the Virgo gEs M49 and M87.  In M49, we have a 
total of 144 confirmed clusters out to 8 arcmin radius
($\sim$ 6 R$_{eff}$ or 35 kpc).
We find that the blue (metal-poor) clusters have both a higher
velocity dispersion and rotation than the red (metal-poor) clusters. 
For the metal-rich population we place an upper limit of 
(v/$\sigma$)$_{proj}$ $<$ 0.34 at 99\% confidence. 
We calculate the velocity dispersion as a function
of radius, and show that this is consistent with isotropic cluster orbits
and the M49 mass distribution determined from X-ray
data.  For M87, we combine new CFHT data with previous data 
to obtain a total sample of 278 clusters out 
to 10 arcmin radius ($\sim$ 45 kpc).  We
find a similar global rotation for the metal-poor and metal-rich clusters
of 160$-$170 km/sec.  Beyond $\simeq$ 2 R$_{eff}$ (15 kpc),
both the metal-poor and metal-rich clusters appear to rotate about the 
photometric {\it minor} axis.  The combined cluster sample is consistent
with isotropic orbits, but when considered separately, the metal-poor clusters
show significant {\it tangential} bias of $\beta_{cl}$ $\simeq$ $-$0.4, while
the metal-rich clusters show a {\it radial} bias with $\beta_{cl}$ $\simeq$ $+$0.4. 
In both galaxies, the metal-rich and metal-poor clusters share
different kinematics, but there is no clear preference for any one formation
scenario.

\end{abstract}

\section{Introduction}

There is still considerable controversy about how and when elliptical
galaxies form.  As survivors from the earliest epochs of galaxy formation,
globular clusters (GCs) are incredibly useful probes of their host galaxies.
Of great recent interest has been the finding of
{\it bimodality} in the
GC colour distributions in many elliptical galaxies.  Several models have
been put forward to explain this bimodality, including spiral-spiral
mergers (Schweizer 1987; Ashman \& Zepf 1992), multi-phase collapse
(Forbes, Brodie, \& Grillmair 1997), and hierarchical accretion 
(C\^ot\'e, Marzke, \& West 1998).

The {\it kinematics} and {\it ages} of metal-poor and metal-rich GCs 
around ellipticals will be key to distinguishing between these different
scenarios, hence the importance of cluster spectroscopy.  However,
this spectroscopy is difficult, given the faintness of GCs 
in even the nearest ellipticals.  It has only been with the advent of 
multi-object spectrographs on 4m and 8$-$10m telescopes that real progress has
been made in this area.
However, the sample of ellipticals with
signficant numbers of GC spectra is still very small, limited to
giant elliptical (gE) galaxies such as Centaurus A, NGC 1399, and
M49 and M87 in Virgo.
Here we present recent multi-slit spectroscopy of GCs in 
the latter two galaxies.

\section{M49 Globular Clusters}

{\it With K. Ashman, M. Beasley, D. Geisler, D. Hanes, R. Sharples,
and S. Zepf.  
See Sharples et al. (1998) and Zepf et al. (2000) for further
details of the M49 GC kinematics,  
and Beasley et al. (2000) regarding the GC 
ages
and metallicities.}

\subsection{Sample Selection and Observations}

Object selection was based on Washington photometry of M49 cluster candidates
from Geisler, Lee, \& Kim (1996). 
To improve the GC yield,
we imposed magnitude and colour cuts:  19.5 $<$ V $<$ 22.5, 
0.5 $<$ C$-$T$_1$ $<$ 2.2. 
Cluster spectra were obtained at the WHT with LDSS in 1994, and at the
CFHT with MOS in 1998.  Spectra covered the region from 
3800$-$6000 \AA\ with a resolution of 3$-$6 \AA, giving 
velocities good to 50$-$100 km/sec; exposure times were 3$-$4 hours
per mask.
The combined cluster sample from the two runs (also
including some GCs from Mould et al. 1990) is 144; of these, 
93 are blue/metal-poor and 51 are red/metal-rich (dividing
at C$-$T$_1$ = 1.625, or [Fe/H] $\sim$ $-$1.6).  We
suffered only 15\% contamination from non-clusters, and our overall
completness was 65\%.  Our sample extends out in radius to $\sim$ 8 arcmin
($\sim$ 6 R$_{eff}$, or $\sim$ 35 kpc for D=15 Mpc).  Cluster velocities
were obtained from cross-correlation with radial velocity 
stars.

\subsection{Results}

\subsubsection{Kinematics}

Figure 1(a) shows the GC rotation and velocity
dispersion radial profiles, for the total sample, and for the blue and red GCs 
separately.  These profiles were obtained by smoothing with a Gaussian
kernel with $\sigma$ = 100\arcsec\ . 
The blue GCs have a significantly larger velocity dispersion
than the red GCs (300$-$350 km/sec compared to $\sim$ 200 km/sec),
as already found by Sharples et al. (1998).  Figure 1(a) shows that 
the M49 GCs have a slowly declining velocity dispersion profile, 
seen in both the full and blue/red samples.  There
is little evidence for rotation in the red GCs (but some evidence that the
rotation increases outwards), while the rotation is roughly constant 
for the blue GCs.  At all radii, the rotation is much smaller than the
dispersion, and for the red GCs we can set an upper limit of
(v/$\sigma$)$_{proj}$ $<$ 0.34 at 99\% confidence; thus, rotation is not
dynamically important for the M49 GCs.


\begin{figure}[h]
\centerline{\hbox{
\psfig{figure=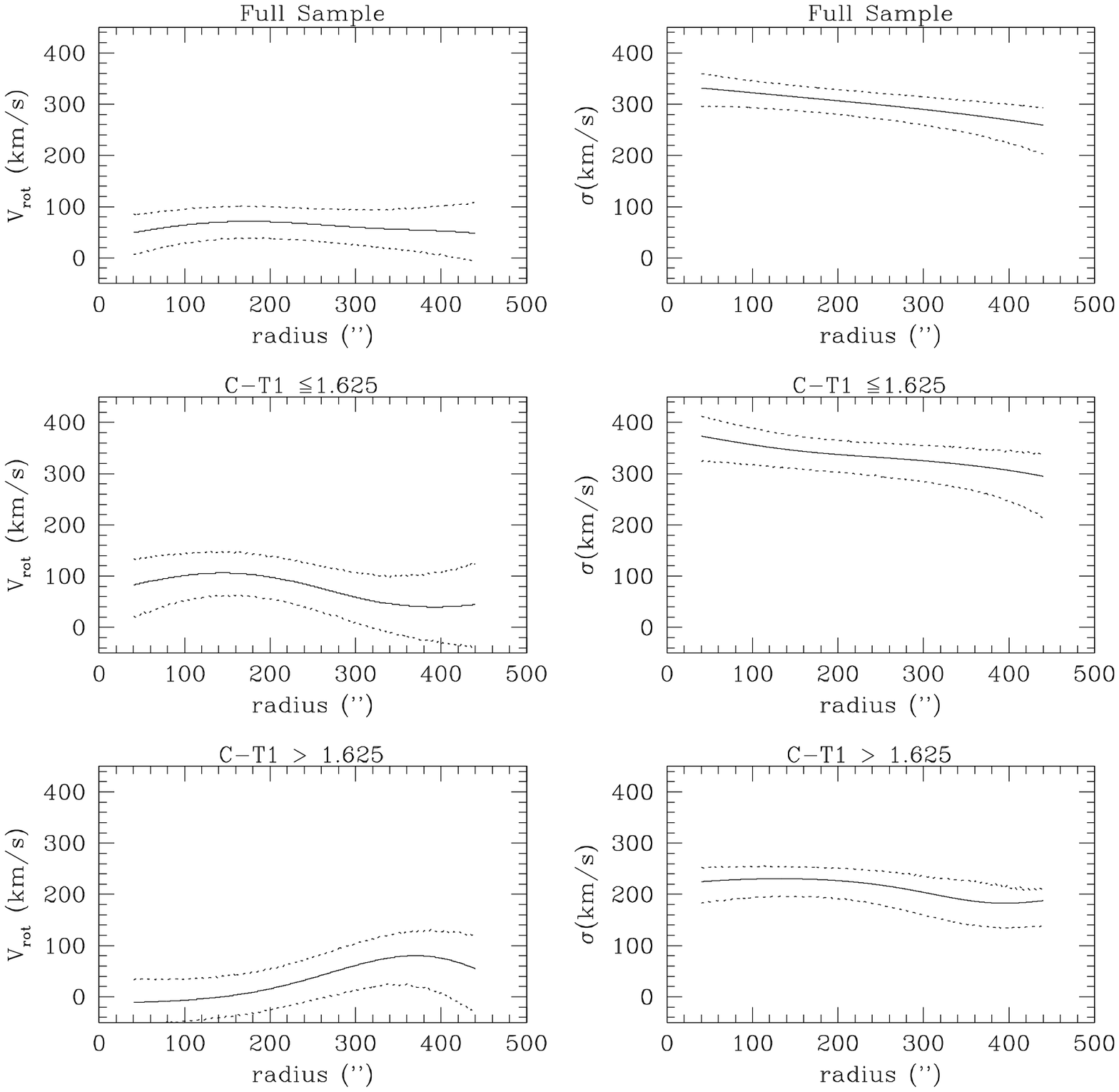,width=5cm,height=5cm}
\psfig{figure=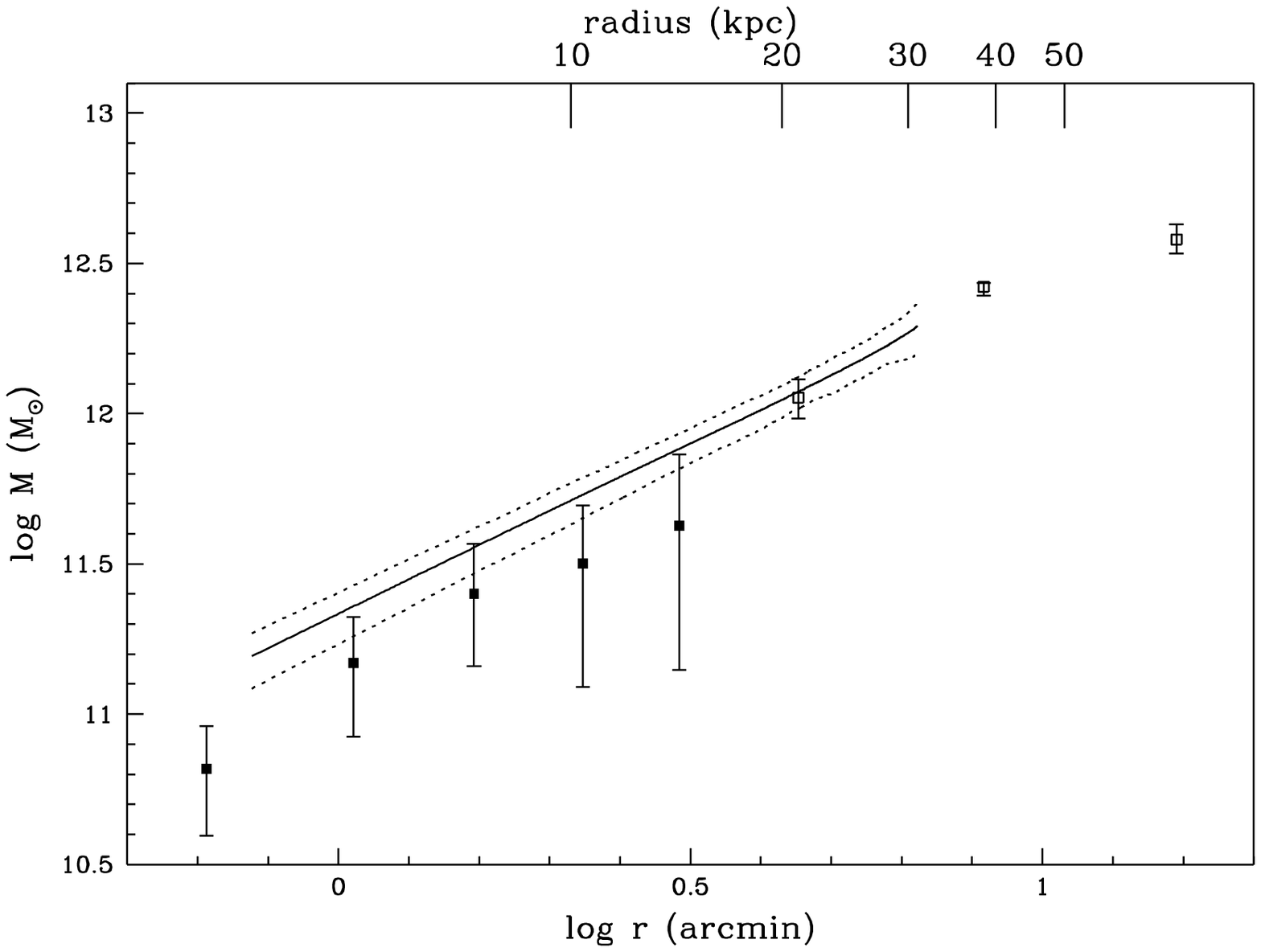,width=5cm,height=5cm}
}}
\caption{{\bf (a) (left)}: Smoothed rotation and velocity dispersion profiles 
for the
M49 GCs.  {\it Top panels}: combined sample;
{\it Middle panels}: blue (metal-poor) GCs; {\it Bottom
panels}: red (metal-rich) GCs.  
Dotted lines show the 1$\sigma$ bootstrapped uncertainties. \\
{\bf (b) (right)}: M49 mass vs. radius.
{\it Solid line}: mass determined from the cluster radial velocities,
assuming isotropic orbits; {\it Dotted lines}: 1$\sigma$ bootstrapped
uncertainties; {\it Points with error bars}: masses from ROSAT X-ray data (Irwin
\& Sarazin 1996).}
\end{figure}

\subsubsection{M/L Ratio and Orbital Anisotropy}

We use the velocity dispersion and luminosity density profiles, together with
the Jeans equation, to estimate the
M49 mass distribution; we assume isotropic orbits
and spherical symmetry.
Figure 1(b) shows the M49 mass vs. radius; the lines 
represent our cluster data,
and the points with errorbars are the masses inferred
from ROSAT X-ray data (Irwin \& Sarazin 1996). 
There is reasonable
agreement between the two mass determinations, suggesting that both are 
roughly correct, and that our assumption of isotropic cluster orbits isn't
too far wrong.  
We find a total M49 mass in excess of 10$^{12}$ M$_\odot$, and a
M/L ratio at least five times greater at
30 kpc than at a few kpc, 
confirming the existence of a substantial dark matter halo in 
M49.

\section{M87 Globular Clusters}

{\it with P. C\^ot\'e, D. Geisler, D. Hanes, G. Harris,
J. Hesser, D. McLaughlin, and D. Merritt.  See Hanes
et al. (2001) and C\^ot\'e et al. (2001) for further details.}

\subsection{Observations and Final Sample}

Object selection was based on Washington C, T$_1$ imaging
(Geisler, Lee, \& Kim 2001), with 0.8 $\leq$ C$-$T$_1$ $\leq$ 2.35.  Spectra
were obtained at the CFHT in 1996 with MOS, with a
dispersion of 3.6 \AA/pixel.
We used a filter centered at 5100 \AA\
with a bandwidth of 1200 \AA\ to allow us to place up to 100 slits 
per mask without
spectral overlap.  Galactic GCs were observed as radial
velocity templates.  We obtained velocities for 109 M87 GCs,
54 of these new measurements.  By combining with previous work
(most importantly, 205 velocities from Cohen \& Ryzhov 1997),
we obtain a final sample of 278 {\it bona-fide} M87 GC 
velocities.  Of these, 161 are classified as ``blue" (C$-$T$_1$ $<$
1.4) or metal-poor ([Fe/H] $<$ $-$1.15) and 117 as ``red" or metal-rich
(C$-$T$_1$ $>$ 1.4, [Fe/H] $>$ $-$1.15). 

\subsection{Results}

\subsubsection{Cluster Kinematics}

We investigate the {\it global} 
rotation and velocity dispersion of the M87 GCS by fitting
our velocities with the function 
$v_p(\Theta) = v_{sys} + (\Omega R){\rm sin}(\Theta-\Theta_0)$,
where $\Theta_0$ is the GCS rotation axis, and $\Omega$R is the
rotation amplitude.
When we average over all radii, we find similar rotation, dispersion, and
position angles for the full sample, and for the blue and red clusters
separately:  $\Omega$R $\sim$ 160$-$170 km/sec, a dispersion of
360$-$400 km/sec, and a position angle between 60$-$75\deg.

When we look at the kinematics as a function of radius, however,
it starts to get interesting. 
Figure 2(a) shows the cluster rotation amplitude vs. radius for 
the total sample, and for the red and blue clusters separately, while
Figure 2(b)
shows the cluster best-fit position angle vs. radius. 
These (smoothed) 
curves were obtained
by sliding a bin of radial width $\Delta$R= 90\arcsec\ ($\simeq$
6.5 kpc) through the cluster datapoints, and solving for the kinematics
in each bin.  From Figure 2(a), we see that the rotation is roughly
constant with radius, but there is some evidence that the rotation 
increases at larger radius, especially for the blue 
GCs.  Figure 2(b) shows that, at large radius, the blue GCs  
rotate conventionally around the minor axis of M87, but for R $\leq$
16$-$18 kpc, they appear to rotate about the {\it major} axis!  Similar
behavior is seen in the red
GCs, though not as pronounced. 
This flip in the rotation axis may be evidence for a past major merger. 

\begin{figure}[h]
\centerline{\hbox{
\psfig{figure=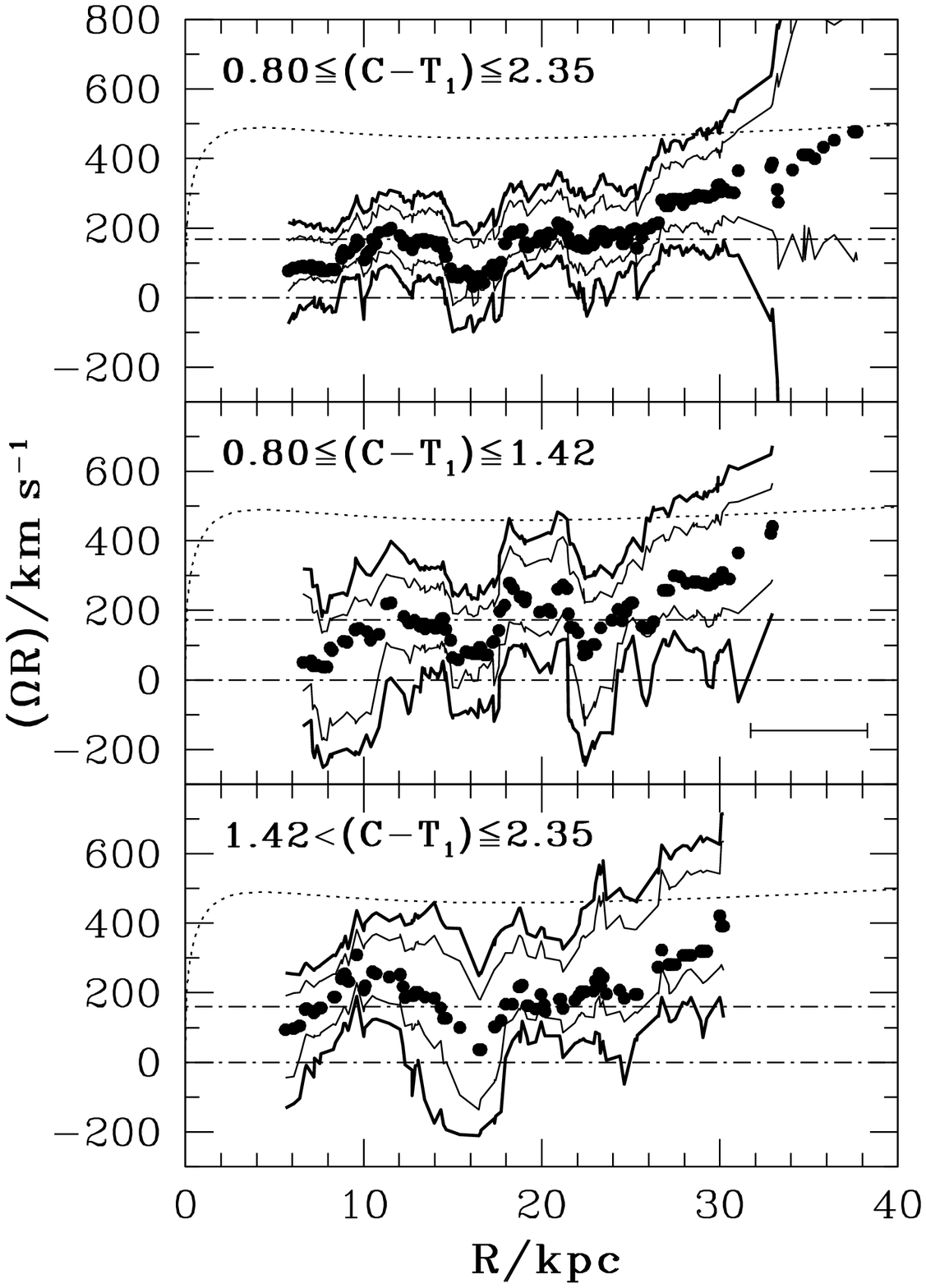,width=5cm,height=5cm}
\psfig{figure=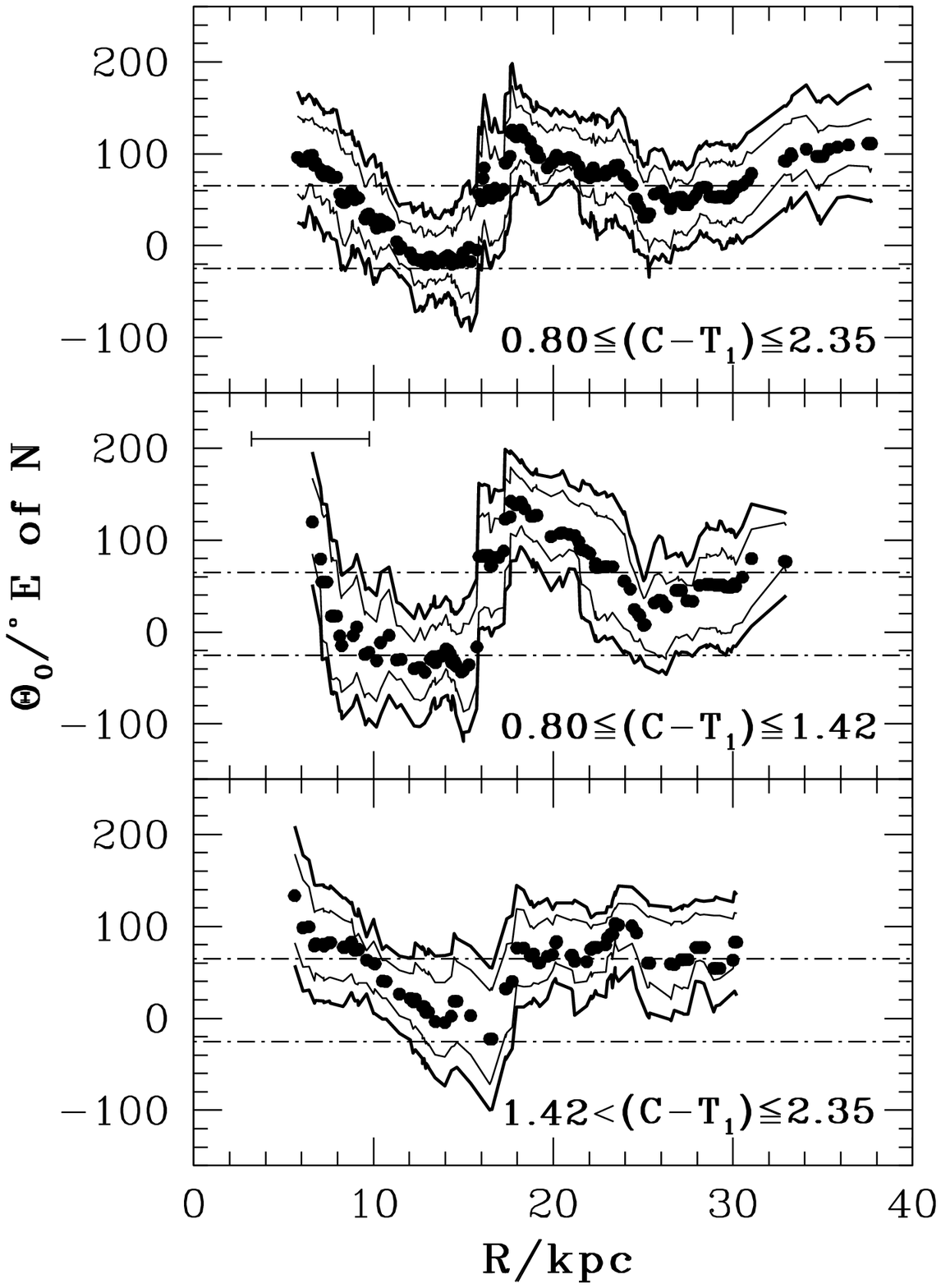,width=5cm,height=5cm}
}}
\caption{ {\bf (a) (left)}: Projected rotation velocity vs galactocentric
radius for the M87 GCS (filled circles), with 68\% and 90\% confidence
bands (thin and thick solid lines).   The smooth, dotted curves
represent the circular velocity of the M87/Virgo potential from
the model in McLaughlin 1999.  {\it Upper
panel}: combined sample; {\it Middle panel}: metal-poor
GCs; {\it Bottom panel}: metal-rich GCs. \\
{\bf (b) (right)}: Same as Figure 2(a), but now plotting the
projected azimuth of the rotation axis of the M87 GCS
vs galactocentric radius.  The lower and upper horizontal lines show the
position angles of the M87 major and minor axes.}
\end{figure}

\subsubsection{Dynamical Models and Cluster Orbits}

Our dynamical approach is to {\it assume} a total M87 mass profile, 
as given by McLaughlin (1999). 
We plug this mass profile, the deprojected
globular cluster surface density profile, and the velocity anisotropy
$\beta_{cl}(r) \equiv 1 - \sigma^2_\theta/\sigma^2_r$ into the Jeans
equation and solve for the radial velocity dispersion $\sigma_r$(r).
A deprojection then gives the projected velocity dispersion profile
$\sigma_p$(r), which can then be compared with the observed profiles.
Thus, we input various values of $\beta$ [specifically $\beta$ = $-$99 (strong
tangential bias), $-$0.4 (moderate tangential bias), 0 (isotropic),
+0.4 (moderate radial bias), and 0.99 (strong radial bias)], and see
which value fits the data best. 

Figures 3(a-c) show the results of this, for the combined sample, and 
for the metal-poor and metal-rich GCs separately.  From Figure 3(a),
the M87 GCS {\it as a whole} has an almost perfectly isotropic velocity
ellipsoid.  However, the metal-poor GCs appear to have a modest
{\it tangential} bias, with $\beta_{cl}$ $\sim$ $-$0.4 at small radii,
while the metal-rich GCs appear to have a slight {\it radial} bias
of roughly the same magnitude, $\beta_{cl}$ $\sim$ +0.4.  Thus, the
metal-poor and metal-rich GCs appear to have quite different orbits.

\begin{figure}[h]
\centerline{\hbox{
\psfig{figure=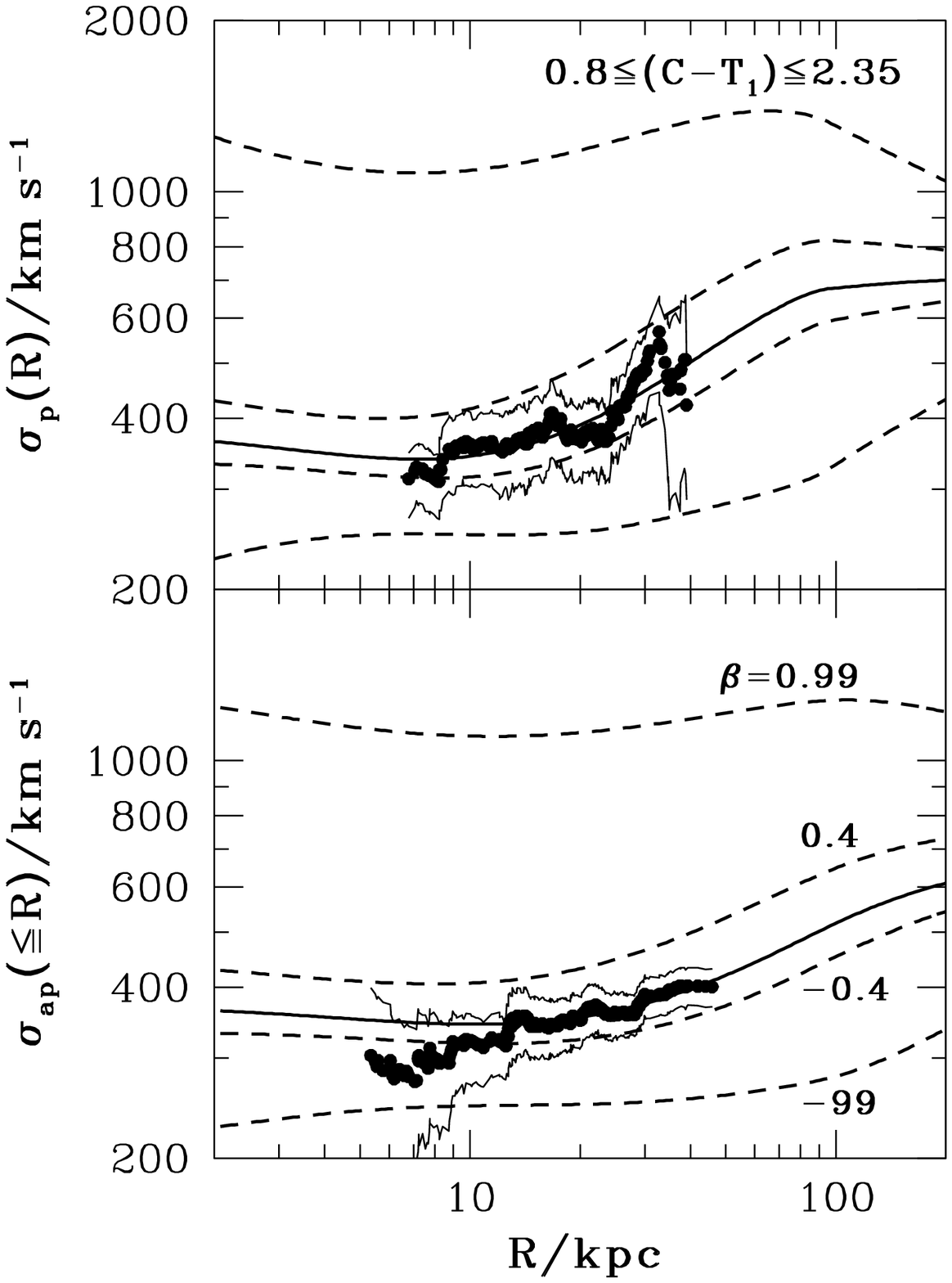,width=5cm,height=5cm}
\psfig{figure=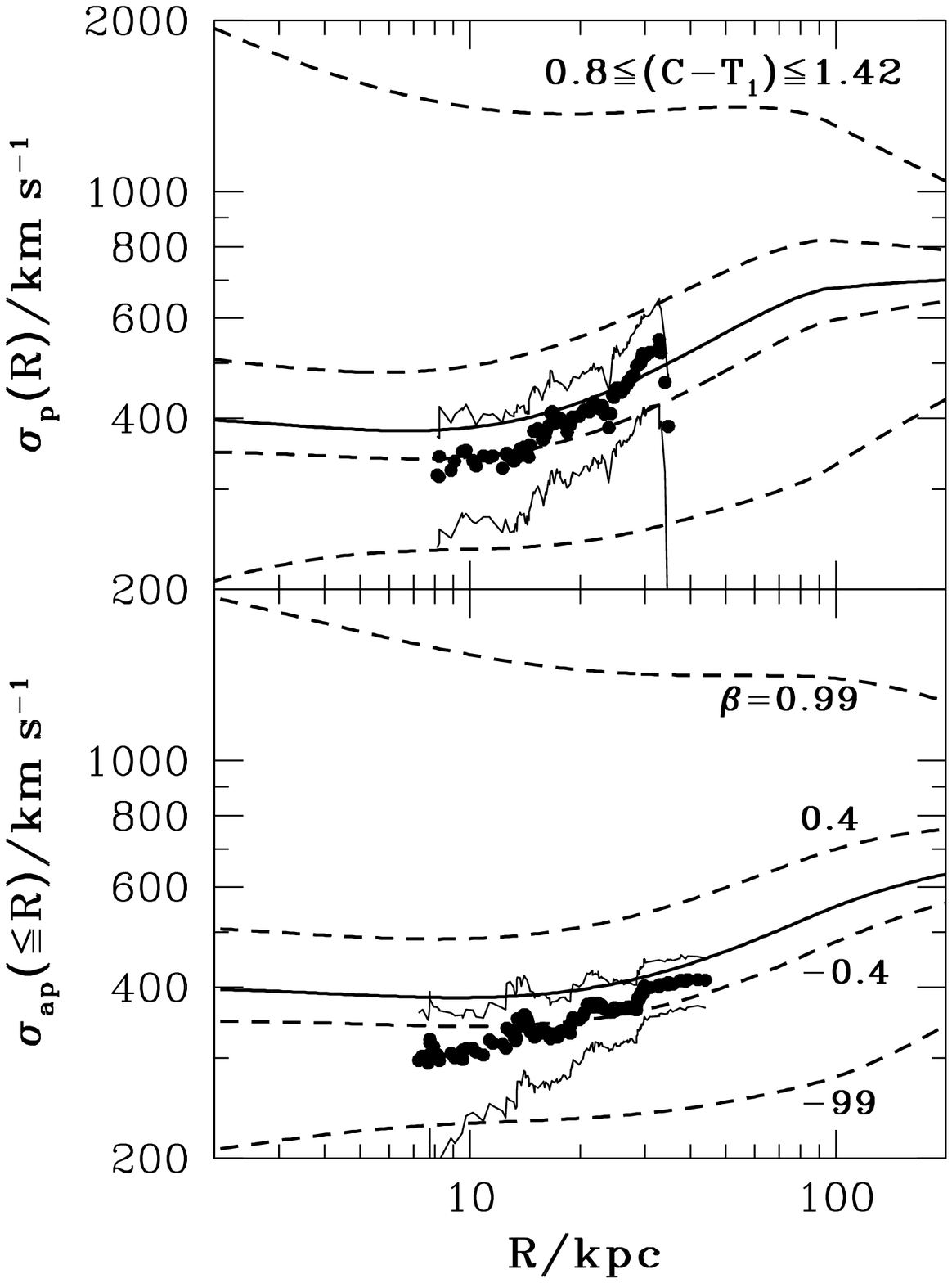,width=5cm,height=5cm}
\psfig{figure=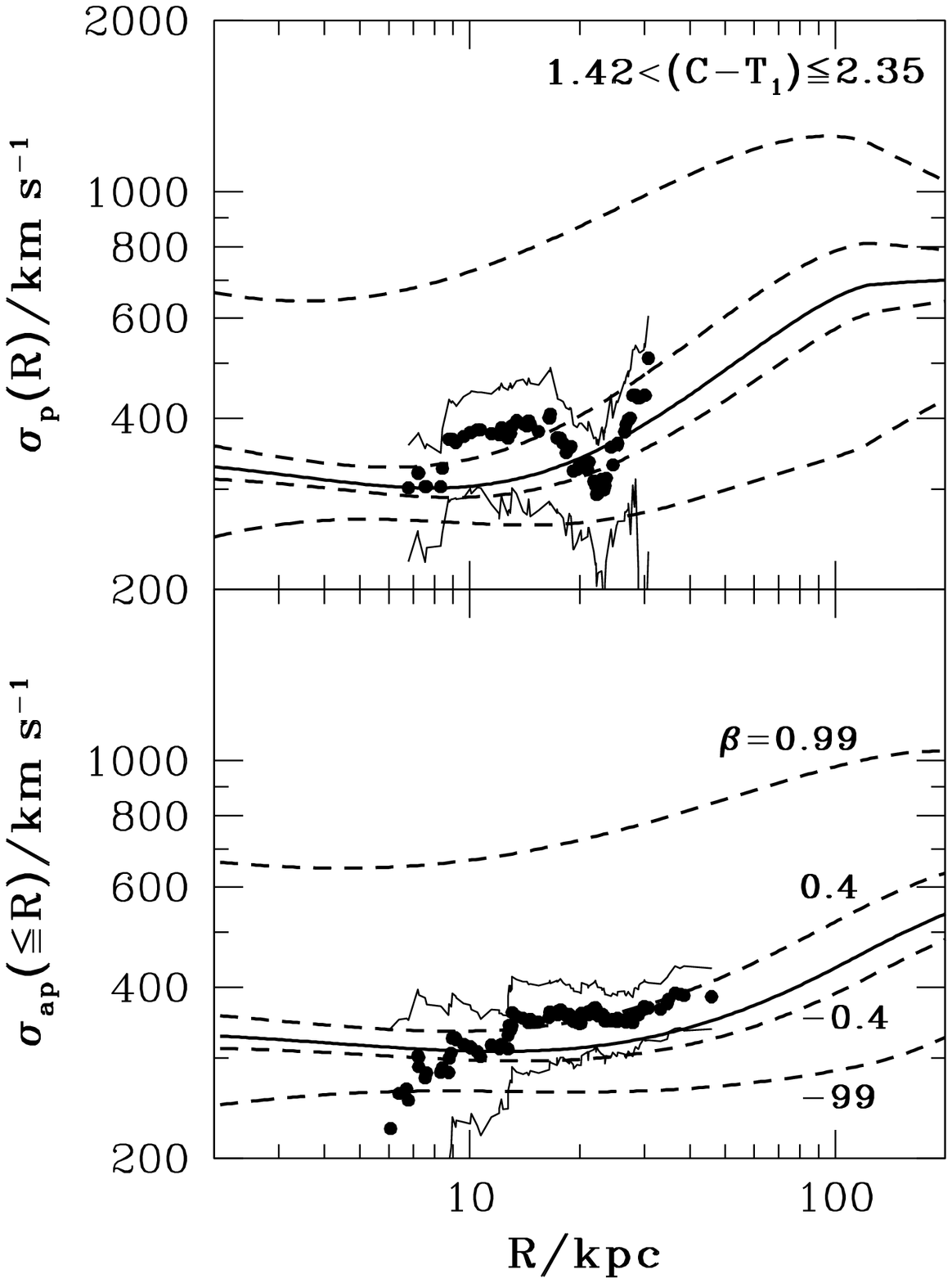,width=5cm,height=5cm}
}}
\caption{ {\bf (a) (left)}: {\it Upper Panel}: Velocity dispersion profile for
the combined sample of M87 GCs.  {\it Points}: smoothed data
(the thin solid curves are 90\% confidence intervals); {\it Thick
solid line}: {\it predicted} velocity dispersion profiles for 
isotropic orbits ($\beta$= 0); {\it Dashed lines}: $\beta$=0.99,
0.4, $-$0.4, and $-$99, from top to bottom.  {\it Lower Panel}: similar to
upper panel, except now for the {\it aperture} velocity dispersion profile.\\
{\bf (b) (middle)}: Same as Fig 3(a), except for the metal-poor GCs.\\
{\bf (c) (right)}: Same as Fig 3(a), except for the metal-rich GCs.}
\end{figure}

\section{Discussion: Comparison of M49 and M87}

What can we learn by comparing the M49 and M87 GCSs?  

\begin{itemize}

\item In M49, there is
little or no rotation in the red GCs, and $\sim$ 100 km/sec rotation
for the blue GCs.  In M87, there is a similar rotation of $\sim$ 170
km/sec for both red and blue GCs.

\item In M49, there is little or no evidence that the rotation or
velocity dispersion is rising with radius, whereas in M87 there is a 
suggestion that both the rotation and dispersion are increasing outwards.
This is most likely because M87 is at the center of Virgo, and the 
M87 GCs are tracing the transition to the Virgo cluster potential.
See C\^ot\'e et al. (2001) for more discussion on this point.

\item In M49, V/$\sigma$ $\sim$ 0.3/0.1 for the blue/red GCs,
while in M87 V/$\sigma$ $\sim$ 0.45 for both subpopulations.

\item The {\it combined} samples of GCs in M49 and M87 are both consistent
with isotropy.  In M87, the blue GCs appear to have a tangential bias,
while the red GCs have a radial bias.

\end{itemize}

Thus, in both galaxies, the metal-poor and metal-rich GCs have different
kinematics (rotation, velocity dispersion, and orbital anisotropy), but these
differences are not consistent between the two galaxies.  It is difficult
to think of a single formation model that can explain the properties (kinematic
and other) of 
both GCSs.  There is obviously a huge need for further kinematical samples
in a wider range of galaxies.  Luckily, we will soon have fabulous new
multi-slit spectrographs on 8m telescopes: GMOS on Gemini and VIMOS 
on the VLT.  The next few years will be very exciting.

\acknowledgements

I'd like to thank my collaborators for allowing me to discuss our work at
this conference, and for the pleasure of working with them. I'd
also like to thank the organizers, particularly Doug and Eva, for arranging
such a fantastic conference, and for financial support.  It was a wonderful
week.



\begin{references}

\reference Ashman, K.M., \& Zepf, S.E. 1992, ApJ, 384, 50

\reference Beasley, M.A., Sharples, R.M., Bridges, T.J., Hanes, D.A.,
Zepf, S.E., Ashman, K.M., \& Geisler, D. 2000, MNRAS, 318, 1249

\reference Cohen, J.G., \& Ryzhov, A. 1997, ApJ, 486, 230

\reference C\^ot\'e, P., Marzke, R.O., \& West, M.J. 1998, 501, 554

\reference C\^ot\'e, P., McLaughlin, D.E., Hanes, D.A., Bridges, T.J.,
Geisler, D., Merritt, D., Hesser, J.E., Harris, G.L.H., \&
Lee, M.G. 2001, ApJ, in press (astroph 0106005)

\reference Forbes, D.A., Brodie, J.P., \& Grillmair, C.J. 1997, AJ, 113, 1652 

\reference Geisler, D., Lee, M.G., \& Kim, E. 1996, AJ, 111, 1529 

\reference Geisler, D., Lee, M.G., \& Kim, E.  2001, in preparation

\reference Hanes, D.A., C\^ot\'e, P., Bridges, T.J., McLaughlin, D.E.,
Geisler, D., Harris, G.L.H., Hesser, J.E., \& Lee, M.G. 2001, ApJ, in press
(astroph 0106004)

\reference Irwin, J.A., \& Sarazin, C.L. 1996, ApJ, 471, 683

\reference McLaughlin, D.E. 1999, ApJ, 512, L9

\reference Mould, J.R., Oke, J.B., de Zeeuw, P.T., \& Nemec, J. 1990, AJ, 99, 1823

\reference Schweizer, F. 1987, in {\it Nearly Normal Galaxies}, (New York:
Springer-Verlag), 18

\reference Sharples, R.M., Zepf, S.E., Bridges, T.J., Hanes, D.A.,
Carter, D., Ashman, K.M., \& Geisler, D. 1998, AJ, 115, 2337

\reference Zepf, S.E., Beasley, M.A., Bridges, T.J., Hanes, D.A.,
Sharples, R.M., Ashman, K.M., \& Geisler, D. 2000, AJ, 120, 2928

\end{references}
\end{document}